\documentclass[aps,pre,twocolumn,showpacs,superscriptaddress,preprintnumbers,amsmath,amssymb]{revtex4}
\usepackage{amsmath}
\usepackage{graphicx}
\usepackage{dcolumn}
\usepackage{bm}
\usepackage{float}

\begin{document}
\preprint{\textit{Preprint}}
\title{Bright X-ray source from a laser-driven micro-plasma-waveguide}
\author{Longqing Yi}\thanks{Authors to whom correspondence should be addressed. Electronic mail: yi@uni-duesseldorf.de}
\affiliation{Institut f$\ddot{u}$r Theoretische Physik I, Heinrich-Heine-Universit$\ddot{a}$t D$\ddot{u}$sseldorf, D$\ddot{u}$sseldorf, 40225 Germany}\affiliation{State Key Laboratory of High Field Laser Physics, Shanghai Institute of Optics and Fine Mechanics, Chinese Academy of Sciences, P.O. Box 800-211, Shanghai 201800, China}
\author{Alexander Pukhov}
\affiliation{Institut f$\ddot{u}$r Theoretische Physik I, Heinrich-Heine-Universit$\ddot{a}$t D$\ddot{u}$sseldorf, D$\ddot{u}$sseldorf, 40225 Germany}
\author{Phuc Luu Thanh}
\affiliation{Institut f$\ddot{u}$r Theoretische Physik I, Heinrich-Heine-Universit$\ddot{a}$t D$\ddot{u}$sseldorf, D$\ddot{u}$sseldorf, 40225 Germany}
\author{Baifei Shen}
\affiliation{State Key Laboratory of High Field Laser Physics, Shanghai Institute of Optics and Fine Mechanics, Chinese Academy of Sciences, P.O. Box 800-211, Shanghai 201800, China}
\date{\today}
\newcommand{\mum}{~$\mu$m}
\newcommand{\wcm}{~W/cm$^{2}$}
\newcommand{\gcm}{~g/cm$^{3}$}
\newcommand{\ccm}{~cm$^{-3}$}
\newcommand{\TVM}{~TV/m}
\newcommand{\Carbon}{C$^{6+}$}
\newcommand{\Gold}{Au$^{1+}$}
\newcommand{\Carbontwo}{C$^{2+}$}
\newcommand{\Hydrogen}{H$^{+}$}

\setlength{\belowdisplayskip}{5pt} \setlength{\belowdisplayshortskip}{0pt}
\setlength{\abovedisplayskip}{5pt} \setlength{\abovedisplayshortskip}{0pt}

\begin{abstract}
Owing to the rapid progress in laser technology, very high-contrast femtosecond laser pulses of relativistic intensities become available. These pulses allow for interaction with micro-structured solid-density plasma without destroying the structure by parasitic pre-pulses. This opens a new realm of possibilities for laser interaction with micro- and nano-scales photonic materials at the relativistic intensities. Here we demonstrate, for the first time, that when coupling with a readily available 1.8 Joule laser, a micro-plasma-waveguide (MPW) may serve as a novel compact x-ray source. Electrons are extracted from the walls and form a dense self-organized helical bunch inside the channel. These electrons are efficiently accelerated and wiggled by the waveguide modes in the MPW, which results in a bright, well-collimated emission of hard x-rays in the range of 1$\sim$100 keV.
\end{abstract}
\maketitle
During the last decade, the research on high-brightness compact x-ray sources has made a significant progress while being motivated by many applications in fundamental science, industry, and medicine. Synchrotrons \cite{s1,s2} and free-electron lasers \cite{s3,s4}  can now produce x-ray beams with unprecedented photon flux and brilliance. These large facilities are excellent x-ray sources being booked by a large community of international users. However, these are unique, expensive, and large-scale devices. In parallel $-$ as an alternative approach $-$ interest in laser-plasma based  sources of secondary radiation is growing constantly\cite{s5,s6,s7,s8}. The recent progress in laser technologies and laser-plasma electron acceleration has enabled the development of a compact all-optical Compton source of hard photons \cite{s9,s10} or a laser-plasma synchrotron-like source \cite{s11}. Laser beams with intensities of $10^{21}$ W/cm$^{2}$ are readily available at several facilities worldwide \cite{s12,s13,s14,s15}. The interaction of these ultra-relativistic intense laser pulses with matter is entering a new era of experimentally unexplored regimes, ushering in revolutionary advances in the research of novel x-ray sources.

\begin{figure*}[!t]
\centering
\vspace{-10pt}
\includegraphics[width=13.5cm]{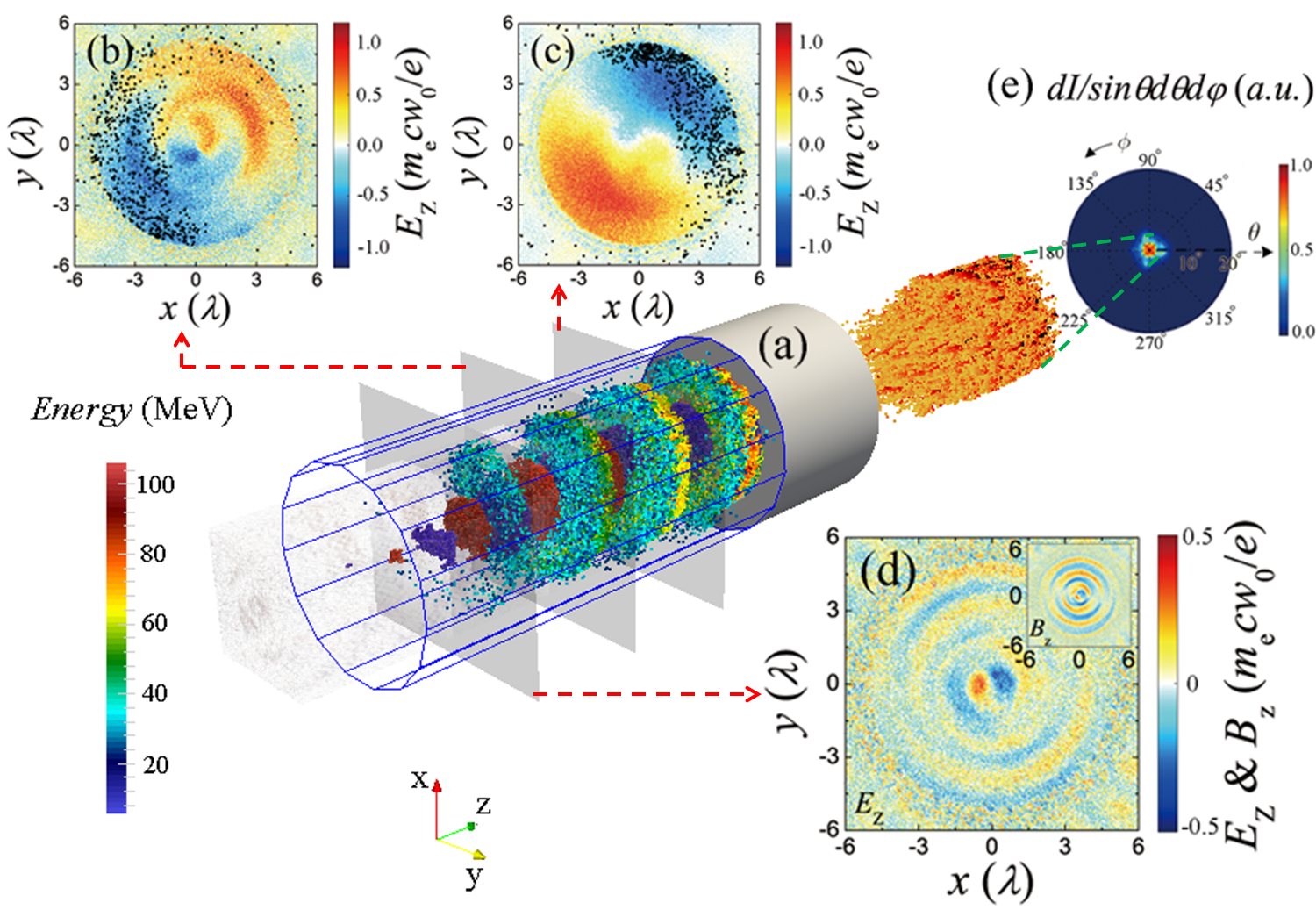}\caption{\label{f1} (Color online.)  (a) Sketch of x-ray production by laser interaction with a MPW. When a circularly polarized laser (blue-red sequence in the middle) propagates inside a MPW (white cylinder and blue wireframe), it extracts electrons out of the walls. These electrons form a helical bunch, which oscillates in the laser field and emits x-ray photons (orange block). The longitudinal electric fields of optical modes excited in the MPW are shown at three cross-sections: middle slice (b), right slice (c), and left slice (d) at $t=50T_{0}$. The colorbar in (a) presents the electron energy at this time. The electron positions at cross-sections (b) and (c) are shown by black dots. On the slice (d), we have shown the longitudinal electric field and magnetic field (inset) observed behind the main pulse, where some high-order optical mode patterns can be seen. (e) Typical radiation pattern and divergence. Here $\theta$ is the angle between the emission and laser propagation direction. The laser pulse has the wavelength $\lambda_0=800~$nm, the pulse energy  is $1.8~$J and the duration is $15~$fs. The MPW radius is $4~\mu$m.}
\vspace{-10pt}
\end{figure*}

For laser-plasma x-ray sources, previous studies have mainly been focused on the radiation from electrons generated by laser-driven wakefield accelerators (LWFA) \cite{s16} or hot electrons produced in laser-foil interactions \cite{s17}. However, the total charge of the electron bunches produced in LWFA is relatively low, which limits the photon yield. On the other hand, in the laser-foil interaction, a high laser-to-photon conversion efficiency ($>1\%$) can be achieved for PW laser pulse \cite{s16}, but the typical radiation divergence is as large as several tens of degrees, making it unsuitable for many applications. Recently, the focus has shifted towards using high contrast laser coupling with structured targets with the aim of increasing laser absorption and the subsequent energy conversion in the secondary radiation \cite{s18,s19}. With recent advances in laser technology and laser pulse cleaning techniques, the laser peak-to-pedestal contrast ratios higher than $10^{10}$ have been achieved using cross-polarized wave generation (XPW) technique \cite{s20}, which allows for interaction with fine plasma structures such as nanoparticles \cite{s21}, snowfakes \cite{s22}, and micro-plasma-waveguides (MPW) \cite{s23}. It has been shown that spatially periodic electron beams with attosecond duration and an over-critical density ($n_{c}=m_{e}\omega_{0}^{2}/4\pi e^{2}$, where $m_{e}$ is the mass of an electron, $e$ is the unit charge, and $\omega_{0}$ is the laser frequency) can be acquired from an intense laser pulse interacting with a MPW \cite{s23,s24,s25}. These electrons are extracted from the boundary surface by the oscillating laser field and accelerated forward via the ponderomotive force. Then, electrons with the proper phase can be further accelerated by the axial component of the laser $E-$field that arises from transverse magnetic (TM) modes in the MPW \cite{s26}. This post-boost mechanism is similar to that in a dielectric laser accelerator \cite{s27}, but since an ultra-intense laser pulse and plasma medium are employed, the acceleration gradient can be as large as several TV/m. Although the maximum energy is limited by phase slippage between the laser phase velocity and electron velocity \cite{s26}, we show that these ultrashort dense energetic electron bunches have the potential to be an excellent source for x-ray emission.

Using multi-dimensional particle-in-cell (PIC) simulations, we report the first numerical observation of bright hard x-ray emission from electrons wiggled by superluminal optical modes in a MPW. A sketch of our simulation set up is shown in Fig.~1(a). When a circularly polarized laser enters the MPW, the electrons from the skin layer of the channel boundary form a self-organized helical bunch owing to the Lorentz force. Electrons with proper phase can be accelerated with a peak acceleration gradient of 4 TV/m as shown in Fig.~1(b) and (c). The map of longitudinal fields in the MPW (Fig.~1(b-d)) shows an interference between different optical modes excited in the MPW. These modes have different phase velocities $v_{p}$, which are all superluminal. As a result, the optical modes overtake the accelerating electron bunches and wiggle them efficiently, which generates well-collimated synchrotron radiations (Fig.~1(e)). The group velocity of the optical modes is $v_{g}=c^{2}/v_{p}$; therefore, the modes with higher order (higher phase velocity) fall behind faster. Some pure high order mode patterns are observed in the rear of simulation box behind the main laser pulse as shown in Fig.~1(d).

Keeping in mind the possible applications of hard x-ray photons, we focus on the angular distribution of radiation and its spectrum, along with the x-ray generation efficiency. We carried out 3D PIC simulations with the code VLPL \cite{s28} to explore the electron dynamics and the synchrotron radiation in the MPW, where a moderately high-intensity laser pulse (intensity $\sim10^{20}$ W/cm$^{2}$, normalized amplitude $a_{0}=15$) is employed. The laser initially has a Gaussian beam profile with temporal FWHM duration duration equal to 15 fs. The waist of the beam is $3.5\lambda_{0}$ ($\lambda_{0}$=0.8 $\mu$m is the laser wavelength). The MPW has a density of $n_{0}=10 n_{c}$, an inner radius of $R_{0}=5\lambda_{0}$ and longitudinal length of 400 $\mu$m. The dimensions of the simulation box are $x\times y\times z=12\lambda_{0}\times 12\lambda_{0}\times 20\lambda_{0}$, and are sampled by 300$\times$300$\times$1000 cells in each direction. The time step is $dt=0.016T_{0}$, where $T_{0}=2.67$ fs is the laser cycle.

Figure~2(a) presents the dependence of radiation intensity on emission angle and photon energy. The x-rays are dominantly in the forward direction, which shows a good agreement with Fig.~1(e). By integrating the photon energy, one obtains the dependence of radiation intensity on emission angle, as presented by the blue curve in Fig.~2(b). Obviously, it has a Gaussian-like distribution with a peak on the laser propagation axis. The root mean square (RMS) opening angle is $\theta_{rms}=17$ mrad. Similarly, the red curve in Fig.~2(c) is obtained by integrating the angle dependence in Fig.~2(a), which shows the photon energy spectrum. One can see a broad frequency band covering the total range from soft x-rays to hard x-rays, with a center frequency around 20 keV.

\begin{figure}[!t]
\vspace{-10pt}
\includegraphics[width=8.5cm]{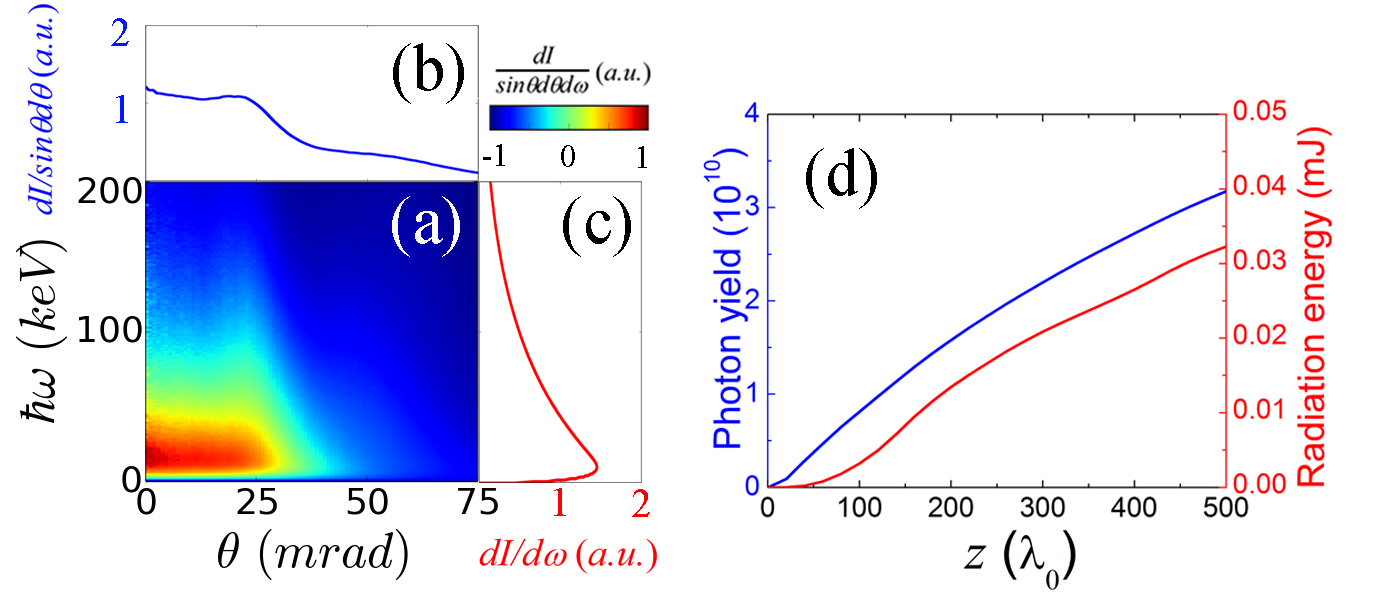}\caption{\label{f1} (Color online.) (a) Angular dependence of the emitted photon energy, and dependence of radiation intensity on (b) angle and (c) photon energy. (d) Total photon yield and radiation energy. Here, $\theta$ is the same as in Fig.~1(e).}
\vspace{-20pt}
\end{figure}

Figure~2(d) presents the photon yield and radiation energy plotted against propagation distance; a linear growth is observed up to $200\lambda_{0}$, which then turns to saturation slowly. The total photon yield is $3 \times 10^{10}$ /shot. Taking a divergence of $17\times17$ mrad$^{2}$, spot size 16$\pi$ $\mu$m$^{2}$ and x-ray pulse duration $\sim$ 15 fs (equal to the duration of laser pulse), the peak brilliance is $1.4\times10^{23}$ photons/s/mm$^{2}$/mrad$^{2}$/0.1$\%$bandwidth, comparable to or even greater than state-of-the-art 3rd-generation synchrotron sources. The total radiated energy is 0.03 mJ, while the incident laser energy is about 1.8 J in the simulation, which results in a reasonable laser-to-photon conversation efficiency of $1.7 \times 10^{-5}$.

In order to obtain a deeper insight into the radiation mechanism, we now analyze the electron dynamics in MPW. Owing to the axially symmetric structure, we consider the cylindrical coordinate system ($z$, $r$, $\phi$), choosing the $z$ axis as the axis of symmetry of the cylindrical MPW; the laser propagates towards the $+z$ direction. The electric and magnetic fields in a MPW can be expressed in terms of two Hertz potentials $\Pi^{e}$ and $\Pi^{m}$ \cite{s29}
\begin{align}
\vspace{-10pt}
\begin{split}
E_{z}=\frac{\partial^{2}\Pi^{e}}{\partial z^{2}}+k^{2}\Pi^{e},
H_{z}=\frac{\partial^{2}\Pi^{h}}{\partial z^{2}}+k^{2}\Pi^{m},
\end{split}
\vspace{-10pt}
\end{align}
\begin{align}
\vspace{-10pt}
\begin{split}
E_{r}=\frac{\partial^{2}\Pi^{e}}{\partial z\partial r}-i\omega\mu_{0}\frac{1}{r}\frac{\partial\Pi^{h}}{\partial\phi},
H_{r}=\frac{\partial^{2}\Pi^{h}}{\partial z\partial r}+i\omega\epsilon_{0}\frac{1}{r}\frac{\partial\Pi^{e}}{\partial\phi},
\end{split}
\vspace{-10pt}
\end{align}
\begin{align}
\vspace{-10pt}
\begin{split}
E_{\phi}=\frac{1}{r}\frac{\partial^{2}\Pi^{e}}{\partial z\partial \phi}+i\omega\mu_{0}\frac{\partial\Pi^{h}}{\partial r},
H_{\phi}=\frac{1}{r}\frac{\partial^{2}\Pi^{h}}{\partial z\partial \phi}-i\omega\epsilon_{0}\frac{\partial\Pi^{e}}{\partial r},
\end{split}
\vspace{-10pt}
\end{align}
where
\begin{align}
\vspace{-10pt}
\begin{split}
\Pi^{e}(z,r,\phi)=\sum_{n,m}a_{n}^{e}J_{n}(T_{m}r)\sin(n\phi)e^{-ik_{z}z},
\end{split}
\vspace{-10pt}
\end{align}
\begin{align}
\vspace{-10pt}
\begin{split}
\Pi^{h}(z,r,\phi)=\sum_{n,m}a_{n}^{h}J_{n}(T_{m}r)\cos(n\phi)e^{-ik_{z}z}.
\end{split}
\vspace{-10pt}
\end{align}
Here, $\epsilon_{0}$ and $\mu_{0}$ are the permittivity and permeability of vacuum, $a_{n}^{e}$ and $a_{n}^{h}$ are coefficients determined by the incident laser amplitude, $k_{z}$ is the longitudinal wave number, $T_{m}=\sqrt{k^{2}-k_{z}^{2}}=x_{m}/R_{0}$ is the transverse wave number, and $x_{m}$ is the m-th root of the eigenvalue equation \cite{s29}
\begin{align}
\vspace{-10pt}
\begin{split}
A(n)[B(n)-2C(n)]+\frac{\omega_{p}^{2}}{\omega^{2}}C(n)K(n)=0,
\end{split}
\vspace{-10pt}
\end{align}
where
\begin{align}
\vspace{-10pt}
\begin{split}
A(n)=J(n)+K(n),
\end{split}
\vspace{-10pt}
\end{align}
\begin{align}
\vspace{-10pt}
\begin{split}
B(n)=A(n)+\frac{\omega_{p}^{2}}{\omega^{2}}[(n/y^{2})-K(n)]
\end{split}
\vspace{-10pt}
\end{align}
\begin{align}
\vspace{-10pt}
\begin{split}
C(n)=n/x^{2}+n/y^{2},
\end{split}
\vspace{-10pt}
\end{align}
\begin{align}
\vspace{-10pt}
\begin{split}
J(n)=\frac{J_{n+1}(x)}{xJ_{n}(x)}, K(n)=\frac{K_{n+1}(y)}{yK_{n}(y)}
\end{split}
\vspace{-10pt}
\end{align}
Here $y = \sqrt{\frac{4\pi^{2}}{\lambda_{p}^{2}}-x^{2}}$, $J_{n}(x)$ is the Bessel function of the first kind, $K_{n}(y)$ is the modified Bessel function of the second kind. Since the most important modes in MPW are TM$_{11}$ and TE$_{11}$, which contain the major part of the incident laser energy, as shown by Fig.~1(b-d). For simplicity, in the following we only take these two modes into consideration. A brief discussion about multi-mode effect will be given after the underlining physics features are addressed. In this situation, the amplitude of Hertz potential in Eq.~(4-5) can be obtained by $a_{1}^{e}=\sqrt{\mu_{0}/\epsilon_{0}}a_{1}^{m}\approx\sqrt{2}a_{0}m_{e}c\omega_{0}/ekT_{1}$.

\begin{figure}[!b]
\vspace{-5pt}
\includegraphics[width=8.5cm]{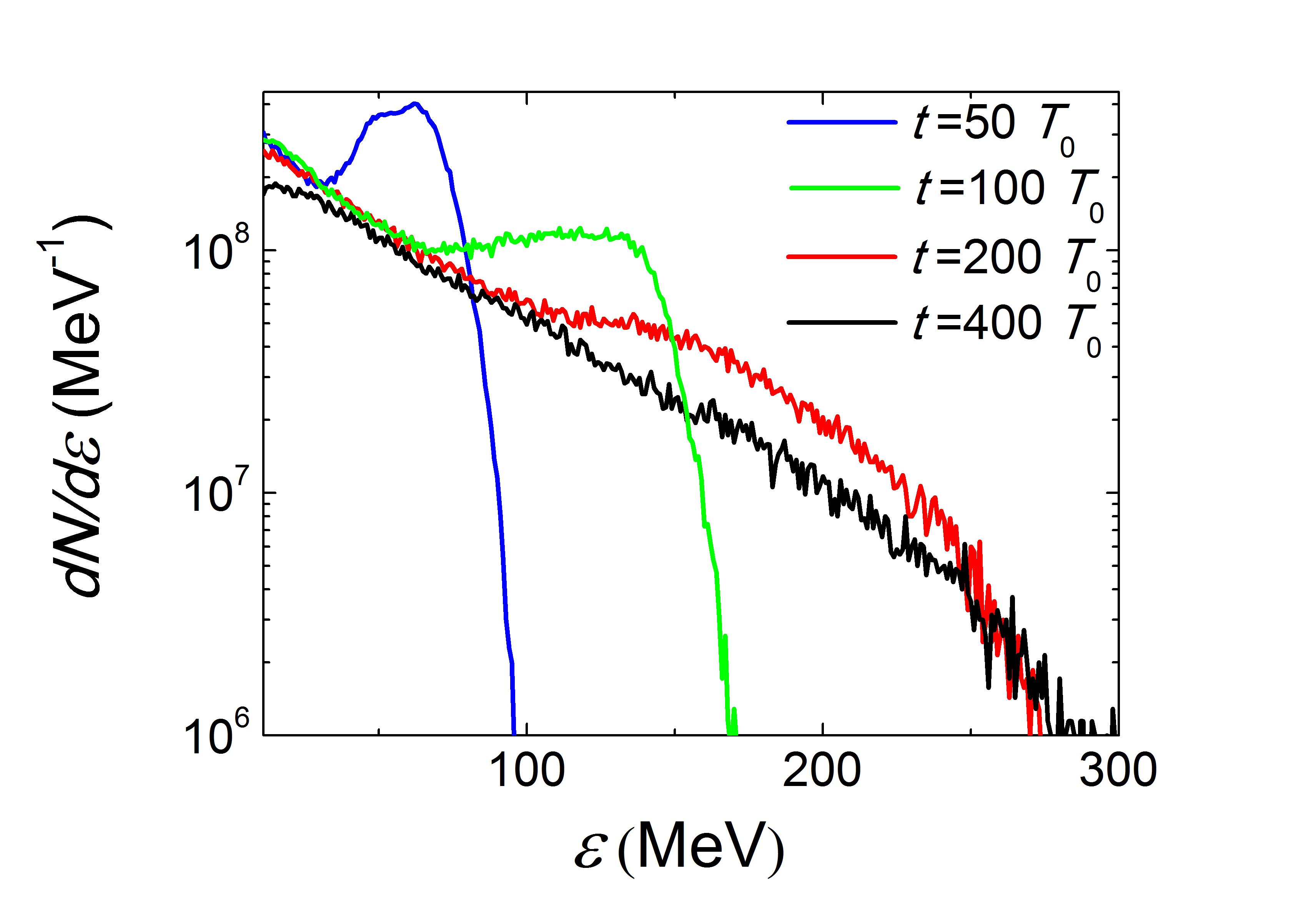}\caption{\label{f1} (Color online.) The electron energy spectrums at different times.}
\vspace{-15pt}
\end{figure}

Since the electron energy plays a key role in the radiation, the acceleration procedure should be discussed first. Since we only interested in the electrons located in the suitable phase (see Fig.~1(b),(c)), the acceleration due to longitudinal component of TM modes can be calculated as
\begin{align}
\vspace{-10pt}
\begin{split}
\varepsilon_{max}\approx e\bar{E}_{z}L_{d}\approx\frac{2\pi^{2}R_{0}a_{0}}{x_{1}\lambda_{0}}mc^{2},
\end{split}
\vspace{-10pt}
\end{align}
where $\bar{E}_{z}\approx TE_{0}/2k$ is the average longitudinal electric field, $E_{0}$ is the incident laser amplitude, $L_{d}=4\pi^{2}R_{0}^{2}/x_{1}^{2}\lambda_{0}$ is the dephasing distance. By submitting $x_{1}=2.5$ we obtain the dephasing distance $L_{d}\approx 150 \lambda_{0}$ and maximum energy as $\varepsilon_{max}\approx300$ MeV, which agrees with our simulation as shown in Fig.~3.

Because the critical frequency of synchrotron radiation proportional to $\gamma^{2}$, the rapid acceleration allows the electrons to radiate at high frequencies in the x-ray or even $\gamma$-ray domain. Meanwhile, in order to study the radiation features, one should also take into account the transverse fields in the MPW. According to Eq.~(2-5), the transverse forces acting on an electron is
\begin{align}
\vspace{-10pt}
\begin{split}
F_{\perp}=\sqrt{F_{r}^{2}+F_{\phi}^{2}}\approx\frac{\sqrt{2}m_{e}c\omega_{0}a_{0}x_{1}^{2}\lambda_{0}^{2}}{8\pi^{2}R_{0}^{2}}J_{0}(Tr)ie^{-ik_{z}z},
\end{split}
\vspace{-10pt}
\end{align}
where we have assumed that the electron velocity is close to the speed of light ($\beta\sim1$) and that the electric magnetic wave propagates mainly along the z axis, i.e. $T_{1}/k\ll1$.

Equations~(12) indicate that the optical modes excited in a MPW exert a net transverse force on the ultra-relativistic co-propagating electrons. The asymmetry in the transverse electric and magnetic fields arises from the transverse component of the wave number $T_{1}$, which induces partial electric (or magnetic) field coupling in the longitudinal direction via refection by the channel wall. Since the radiation power scales with the square of the transverse force ($P\propto \gamma^{2}F_{\bot}^{2}$), the asymmetric structure of the optical modes in a MPW greatly enhances photon emission. Figure~4(a) presents the dependance of radiation power on the laser amplitude $a_{0}$, which shows a rapid growth as $a_{0}$ increases.

When an electron oscillates in a optical mode described by Eqs.~(12), the critical frequency is determined by the Lorentz factor of the electron and maximum value of transverse field, i.e.,
\begin{align}
\vspace{-10pt}
\begin{split}
\omega_{c}=\frac{3\sqrt{2}\bar{\gamma}^{2}x_{1}^{2}\lambda_{0}^{2}a_{0}}{16\pi^{2}R_{0}^{2}}\omega_{0}\approx\frac{3\sqrt{2}\pi^{2}a_{0}^{3}}{16}\omega_{0}.
\end{split}
\vspace{-10pt}
\end{align}
Equation~(13) allows us to estimate the emitted photon energy. By taking $a_{0}=15$, $R_{0}=5\lambda_{0}$, the average Lorentz factor $\bar{\gamma}\approx\gamma_{max}/2\approx300$, we obtain the critical frequency $\hbar\omega_{c}\approx15$ keV, which agrees well with our PIC simulations as seen in Fig.~2(a) and (c). The long tail on the photon spectrum is mainly produced by higher-order modes that are excited in the MPW, which will be discussed later. The average photon energy (ratio of total radiation energy and photon yield) is plotted as a function of laser amplitude $a_{0}$ for a fixed channel radius $R_{0}=5\lambda_{0}$ in Fig.~4(a). It is shown that the numerical observation can be fitted by cube scaling, as suggested by Eq.~(13).

The divergence of the radiation is determined by the maximum angle between the electron velocity and propagation axis. The highly collimated photon beam indicates a highly collimated electron bunch, which can be attributed to the exact balance between the magnetic field induced by the surface current and electrostatic field due to charge separation \cite{s30}. By double integrating Eq.~(12), one obtains the maximum transverse displacement for an near axis electron ($J_{0}(T_{1}r)\approx1$)
\begin{align}
\vspace{-10pt}
\begin{split}
r_{max}\approx\frac{4\sqrt{2}\pi R_{0}^{2}a_{0}}{x_{1}^{2}\lambda_{0}\bar{\gamma}}.
\end{split}
\vspace{-10pt}
\end{align}
Therefore, the typical opening angle of the radiation cone can be calculated as
\begin{align}
\vspace{-10pt}
\begin{split}
\Psi=\frac{2\pi r_{max}}{\lambda_{u}}\approx\frac{\sqrt{2}a_{0}}{\bar{\gamma}}\approx\frac{\sqrt{2}x_{1}\lambda_{0}}{\pi^{2} R_{0}}.
\end{split}
\vspace{-10pt}
\end{align}
where $\lambda_{u}\approx8\pi^{2}R_{0}^{2}/x_{1}^{2}\lambda_{0}$ is the wiggler spatial period. Equation~(15) provides an estimate of the radiation divergence of 70 mrad, which is in rough agreement with our numerical observations. Furthermore, Eq.~(15) indicates that for a fixed MPW, that the radiation divergence does not depend on $a_{0}$, which is confirmed by our PIC simulation as shown in Fig.~4(a). Moreover, from Eq.~(15), one can also obtain the wiggler strength by \cite{s5} $K=\gamma\Psi\approx a_{0}$.

\begin{figure}[!b]
\vspace{-15pt}
\includegraphics[width=8.5cm]{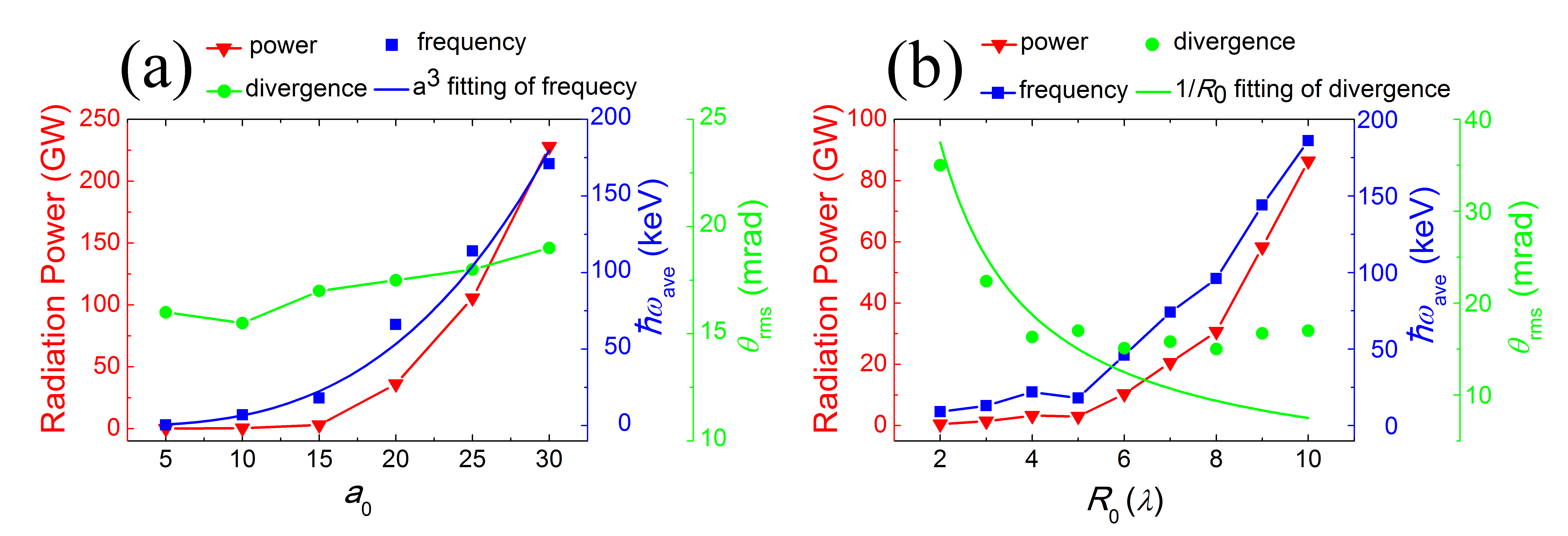}\caption{\label{f1} (Color online.) Dependence of synchrotron radiation power (red triangles), frequency (blue squares) and RMS opening angle (green circles) on (a) laser amplitude and (b) channel radius. The blue curve in (a) shows the $a_{0}^{3}$ fitting of average photon energy, and the green curve in (b) presents the $1/R_{0}$ fitting of divergence.}
\vspace{-10pt}
\end{figure}

We finally discuss the multi-mode effects in the MPW. As a laser pulse is propagating in the MPW, many modes are excited simultaneously, as shown in Fig.~1 (b-d), and the electron motion is subjected to the combined fields of these modes. Typically, higher order modes produce a larger transverse wiggling force owing to the larger value of $x_{m}$, which results in a higher critical frequency and higher radiation power when other parameters are fixed. However, due to the difficulty in obtaining the exact amplitudes and energy loss rate for each optical mode in the MPW, one has to rely on PIC simulations to study the effects induced by high-order optical modes.

Apparently, the multi-mode effects in the MPW is closely related to the radius $R_{0}$, because it decides the highest order of the optical modes excited in a MPW and the efficient interaction length, i.e. $L_{eff}\approx\frac{8\pi^{2}R_{0}^{2}c\tau}{x_{m}^{2}(y_{m}^{2})\lambda_{0}^{2}}$, where $\tau$ is the duration of the laser pulse. As a result, it is an important freedom to adjust the radiation power and photon energy. As it can be seen in Fig.~4(b), the average energy of synchrotron photons and radiation power remain almost unchanged when $R_{0}$ is small and start to grow with $R_{0}$ when it exceeds $6\lambda_{0}$ (laser amplitude is fixed at $a_{0}=15$ and the spot size varies proportional to the channel radius). Simultaneously, the divergence of radiation departs from the $1/R_{0}$ scaling predicted by Eq.~(15). This indicates the excitation of higher optical modes so that the transverse force cannot be estimated by considering the TM$_{11}$ and TE$_{11}$ modes alone. Our numerical results show when increasing the MPW radius, the radiation power is enhanced, and the photon energy increases, while the RMS opening angle decreases and saturates at around 15 mrad as presented in Fig.~4(b).

In conclusion, the interaction of electrons with electromagnetic waves in a MPW has been studied through 3D PIC simulations and theoretical analysis. The MPW can modulate femtosecond relativistic laser pulses at the micro scale, and the proposed novel mechanism has the potential to accelerate electrons to several hundreds of MeV and efficiently wiggle them, resulting in the emission of hard photons with high brightness, broad band width and very low divergence. The underlying physics is discovered, which can be attributed to the fierce acceleration via longitudinal electric field of TM modes, as well as the asymmetric structure of transverse optical mode components. We also discussed the effects induced by high-order optical modes, which have the ability of enhancing the radiation power and photon energy. These effects become increasingly important as the channel radius increases. The emitted tunable brilliant x-rays and $\gamma$-rays have promising features and might have diverse applications in imaging, medical treatment, isotope production, following chemical processes, and nuclear physics.

We thank Dr. L. L. Ji  and Dr. J. Farmer for fruitful discussions and language editing. This work is supported in parts by EUCARD-2, Deutsche Forschungsgemeinschaft SFB TR 18 and National Natural Science Foundation of China (No.11505262, No.11125526, and No.11335013).

\end{document}